\documentstyle[referee]{l-aa}
 
\begin{document}
 
   \thesaurus{01
	      (08.05.3; 08.08.2; 08.22.3; 11.04.1; 11.13.1)  
	     }
   \title{The distance modulus of the Large Magellanic Cloud:}
   \subtitle{Constraints from RR Lyrae pulsation properties}

   \author{M.~Catelan}     
 
   \institute{Instituto Astron\^{o}mico e Geof\'\i sico, 
	      Universidade de S\~{a}o
	      Paulo, C.P. 9638, CEP 01065-970, S\~{a}o Paulo, SP, Brasil \\
	      e-mail: marcio@vax.iagusp.usp.br}
 
   \date{Received; accepted}
   \maketitle
   
   \begin{abstract}

   It has recently been suggested that the discrepancy between the 
   ``long" and ``short" distance
   moduli of the Large Magellanic Cloud (LMC), as inferred from the properties 
   of the Cepheid and RR Lyrae variables, respectively,
   might be due to the action of ``third
   parameters" between the Galaxy and the LMC, which would make the RR Lyraes
   in the old LMC globular clusters brighter than their Galactic counterparts 
   by $\simeq 0.3\, \mbox{mag}$. Through analysis of the RR Lyrae pulsation 
   properties, we show that this idea is not supported by the available data.
   A satisfactory explanation of the problem has yet to be found.
   	 
     \keywords{Stars: evolution -- Stars: horizontal branch -- Stars:
	       variables: other -- Galaxies: distances and redshifts -- 
	       {\em (Galaxies:)} Magellanic Clouds
	       }
   \end{abstract}
 
\section{Introduction}

\noindent In an important {\em Letter}, Walker (1992b) has called attention to 
what appeared to be a fundamental problem with our knowledge of the properties
of the RR Lyrae stars: the
Baade-Wesselink (BW) calibration of the RR Lyrae absolute magnitude-metallicity
relation (e.g., Carney et al. 1992; see also Storm et al. 1994 for a recent
discussion and a comparison between field and cluster results) 
was shown by him to give a distance modulus
for the Large Magellanic Cloud (LMC) that is substantially shorter than 
indicated by the LMC Cepheids (e.g., Laney \& Stobie 1994) and by the 
properties of the SN1987A circumstellar envelope
(e.g., Crotts et al. 1995). The difference was attributed to a 
problem in the zero point of the $M_V({\rm RR}) - {\rm [Fe/H]}$ relation: RR 
Lyrae stars would thus be brighter by $\simeq 0.3\, \mbox{mag}$ than indicated 
by the BW method. Among the implications of this result stand out a 
reduction in the ages of globular clusters (GCs) and a decrease in the value 
of the Hubble parameter $H_0$ (van den Bergh 1995 and references therein).

Independent evidence that the RR Lyrae variables should
be brighter than suggested by the BW method has been presented by 
Saha et al. (1992), Catelan (1992), Simon \& Clement (1993), Cacciari \& Bruzzi 
(1993), Sandage (1993), Dorman (1993), Fernley (1994), Silbermann \& Smith
(1995), etc. Castellani \& De Santis (1994) have shown that 
the BW luminosities
cannot be reconciled with the standard models for evolution on the 
horizontal branch (HB), unless the helium abundance $Y$ is lower than 20\%
by mass, and questioned the accuracy of BW analyses.
Bono \& Stellingwerf (1994), Bono et al. (1994) and Fernley (1994) have 
similarly argued that some of the basic assumptions of the BW method may 
be in error. On the other hand, the
latest analyses of statistical parallaxes of Galactic field halo stars 
(Layden et al. 1994) have reportedly given some support to the BW results.

An explanation different from Walker's (1992b) has been proposed by van den 
Bergh (1995), according to whom the suggested discrepancy
between the distance moduli of the LMC that are inferred through analysis of
the Cepheid and RR Lyrae variables may {\em not} necessarily imply
that the BW absolute magnitudes are incorrect, but rather that a ``third
parameter" is acting in the LMC in such a way as to make the LMC RR Lyraes
intrinsically brighter than those in the Galaxy by $\simeq 0.3\, \mbox{mag}$.

Indeed, the old LMC GCs are somewhat shifted toward redder HB types in
the HB morphology$-{\rm [Fe/H]}$ plane. This has often been 
interpreted (e.g., Walker 1992c) as evidence that the old LMC GCs
are younger by a few Gyr than the Galactic globulars. We note, in passing,
that the very metal-poor Galactic GCs which do not have extremely blue
HB types (e.g., M15 and M68) have {\em not} generally been associated to 
a younger component of the Galactic halo (e.g., van den Bergh 1993; Zinn 
1993), as would be required in the age interpretation of the second-parameter
phenomenon.

In the present {\em Letter}, we submit van den Bergh's (1995) suggestion
to the critical analysis that is enabled by the extensive surveys of RR
Lyrae pulsation properties in the LMC GCs by Walker (1989, 1990, 1992a,c).

\section{Expected trends in second-parameter candidates}

\noindent From the location of the old LMC GCs on the HB morphology-metallicity
plane (see, e.g., Fig. 1 in Catelan \& de Freitas Pacheco 1993), several 
possibilities
emerge for the sense of variation in the known second-parameter candidates: a 
{\em younger age}; a {\em smaller amount of mass lost during the red giant 
branch (RGB) phase};
a {\em lower Y}; a {\em lower helium-core mass at the
helium flash} ($M_{\rm c}$); or a {\em higher relative abundance of the
$\alpha$-capture elements}. These trends are well known from studies of the
evolution of HB stars (e.g., Sweigart \& Gross 1976). The trend of variation
of the HB morphology with [$\alpha$/Fe] was inferred from the analysis of
Salaris et al. (1993). Quantitative estimates of the required changes in
these candidates are, unfortunately, difficult to obtain (Catelan \& de 
Freitas Pacheco 1993).

\noindent -- {\em Age}: the primary effect of age variations is upon the masses
attained by HB stars. As far as the absolute magnitudes of the RR Lyrae 
variables are concerned, age changes are essentially irrelevant. Thus, 
the age interpretation of the 
second-parameter phenomenon would not naturally produce brighter HBs in the LMC;

\noindent -- {\em Mass loss on the RGB}: like age, mass loss by stellar winds 
from the envelopes of RGB stars acts only to reduce the expected masses on the 
HB phase, and has but little impact upon RR Lyrae luminosities;

\noindent -- {\em Helium abundance}: a higher $Y$ in HB stars would
produce brighter RR Lyrae variables in the LMC, in comparison
with the Galactic ones, but also bluer HB types;

\noindent -- {\em Helium-core mass}: a higher $M_{\rm c}$ in the LMC, in 
comparison with the Galactic values, might originate, for instance, in higher
stellar rotation rates (see discussion and references in Catelan et al. 1996). 
This would produce brighter RR Lyrae variables, but would also lead to bluer
HB morphologies;

\noindent -- {\em Abundances of the $\alpha$-elements}: since an overabundance
of the $\alpha$-elements may, in a first approximation, be interpreted in terms
of a higher metallicity $Z$ for a given [Fe/H] ratio, it follows that 
a smaller [$\alpha$/Fe] ratio might account for a brighter HB in the LMC.
It may be noted that existing chemical evolution models (cf. Fig. 4
in Matteucci \& Brocato 1990) suggest that, at low metallicities,
[$\alpha$/Fe] may be lower in the LMC than in the Galaxy. Observational element
ratios at the low metallicities that characterize LMC GCs are badly needed.
However, a smaller [$\alpha$/Fe] ratio would also lead to bluer HBs.

To be sure, there are possible combinations of variations in these
parameters that could account for brighter RR Lyraes and redder HB types
simultaneously. It is conceivable, for instance, that a higher $M_{\rm c}$
could lead to a brighter HB in the LMC, {\em provided that} the LMC GCs
were {\em much} younger than their Galactic counterparts (so as to match their
observed HB types). To put more
stringent constraints on variations in the second-parameter
candidates, analysis of the RR Lyrae pulsation properties is necessary.

\section{Constraints from RR Lyrae pulsation properties}

\subsection{Mean pulsation periods}

\noindent Light curves have been obtained by A. Walker for several GCs of the 
LMC. We have compiled data for the RR Lyrae-rich objects
NGC 2257 (Walker 1989), NGC 1841 (Walker 1990), Reticulum (Walker 1992a),
and NGC 1466 (Walker 1992c). This homogeneous database, supplemented by a 
few entries from Nemec et al. (1985) for NGC 2257, will be employed in
the present discussion.

  \begin{table*}
  \caption[]{Pulsation properties of LMC and Galactic GCs}
  \begin{flushleft}
  \begin{tabular}{lllllllll}
  \noalign{\smallskip}
  \hline  
  \noalign{\smallskip}
  & Cluster & [Fe/H] & $(B-R)/(B+V+R)$ & $\langle \log P_{\rm f} \rangle $ & 
  $\langle \log P_{\rm ab} \rangle$ & 
  $\langle \Delta \log P(T_{\rm eff}) \rangle$ & $N_{\rm RR}$ & $N_{\rm ab}$ \\
  \hline
  LMC & Reticulum & $-1.71$ & $-0.04$ & $-0.283$ & $-0.260$ & $+0.022$ & 31 & 
22 \\
  & NGC 2257  & $-1.8 $ & $+0.49$ & $-0.291$ & $-0.245$ & $+0.041$ & 31 & 15 \\
  & NGC 1466  & $-1.85$ & $+0.40$ & $-0.273$ & $-0.234$ & $+0.037$ & 39 & 23 \\
  & NGC 1841  & $-2.11$ & $+0.72$ & $-0.209$ & $-0.172$ & $+0.072$ & 22 & 17 \\
  Galaxy & M3       & $-1.66$ & $+0.08$ & $-0.276$ & $-0.259$ & $+0.025$ & 179 
& 148 \\
  & M15      & $-2.15$ & $+0.72$ & $-0.250$ & $-0.188$ & $+0.060$ &  67 &  29 \\
  \hline
  \end{tabular}
  \end{flushleft}
  \end{table*}

\begin{figure*}
\rule{0.4pt}{14.5cm}
\hfill      \parbox[b]{4.25cm}{\caption{
	     $P - T_{\rm eff}$ plane for RR Lyrae variables in old 
	     LMC GCs. The mean period shifts obtained at fixed $T_{\rm eff}$
	     with respect to the lower envelope of the M3 
	     distribution (bottom panel) are given, together
	     with the cluster [Fe/H] ratios. The lower envelope of the M3
	     distribution is reproduced in each panel for clarity
	    }}%
	 \label{none}%
\end{figure*}

Mean pulsation periods for these GCs can be found in Table 1. Both
mean ``fundamentalized" periods (obtained by scaling the RRc
periods as $\log P = \log P_{\rm c} + 0.13$) and mean RRab Lyrae
periods are given, together with the number of variables employed in
the analysis and the HB morphology. The [Fe/H] ratios were obtained 
from Walker 1992c or (in the cases of Reticulum and NGC 1841) from 
Suntzeff et al. 1992. Mean period shifts over all ab-type RR Lyraes,
obtained at fixed effective temperature with respect to the lower envelope
of the M3 distribution (cf. Fig. 1), are also displayed. For comparison 
purposes, also given are the corresponding values for the Galactic GCs M3 
and M15, for which the mean periods were drawn from Castellani \& Quarta 1987.

From Catelan's (1993) synthetic
HB models for $Z = 4 \times 10^{-4}$ and $\sigma_M = 0.02\, M_{\sun}$ (where
$\sigma_M$ is the mass dispersion on the HB), one finds that 

\begin{equation}
{ {\rm d} \langle \log P_{\rm f} \rangle \over {\rm d} Y} \approx 1.6
\,\,\mbox{ and } \,\,
{ {\rm d} \langle \log L({\rm RR}) \rangle \over {\rm d} Y} \approx 1.8.
\end{equation} 

\noindent Uncertainties in such slope values are typically estimated to be
of order 10\%. In order to produce a $\Delta M_{\rm bol} = 0.3\, \mbox{mag}$, 
this suggests that the helium abundance in the LMC should be larger than in 
the Galaxy by

\begin{equation}
\Delta Y \approx +0.07.
\end{equation} 

\noindent According to Eqs. (1), this would imply an increase in the mean
fundamentalized periods, in comparison with the Galactic values, by

\begin{equation}
\Delta \langle \log P_{\rm f} \rangle \approx +0.11.
\end{equation} 

\noindent Table 1 does seem to rule out such a large difference in the mean
periods.

Similar arguments apply to an increase in $M_{\rm c}$. For instance, from
the Caputo et al. (1987) synthetic HB models for $Y_{\rm MS} = 0.20$ and
$Z = 4 \times 10^{-4}$, one finds that 

\begin{equation}
{ {\rm d} \langle \log P_{\rm f} \rangle \over {\rm d} \Delta M_{\rm c} } 
\approx 2.3
\,\, \mbox{ and } \,\,
{ {\rm d} \langle \log L({\rm RR}) \rangle \over {\rm d} \Delta M_{\rm c}} 
\approx 3.1.
\end{equation} 

\noindent In order to produce a $\Delta M_{\rm bol} = 0.3\, \mbox{mag}$, 
$M_{\rm c}({\rm LMC})$ should thus be larger 
than $M_{\rm c}({\rm Galaxy})$ by

\begin{equation}
\Delta M_{\rm c} \approx +0.04 \, M_{\sun}.
\end{equation} 

\noindent Equations (4) then show that this would imply an increase in the mean
fundamentalized periods by

\begin{equation}
\Delta \langle \log P_{\rm f} \rangle \approx +0.09.
\end{equation} 

\noindent This again appears to be ruled out by the data.

Catelan's (1993) synthetic HB models also suggest that

\begin{equation}
{ {\rm d} \langle \log L({\rm RR}) \rangle \over {\rm d} \log Z} \approx -0.07.
\end{equation} 

\noindent In order to produce a $\Delta M_{\rm bol} = 0.3\, \mbox{mag}$, 
extrapolation of Eq. (7) toward very low metallicities suggests that the 
metal abundance in the LMC (for a given [Fe/H] ratio) should be smaller than 
in the Galaxy by

\begin{equation}
\Delta \log Z \approx -1.7.
\end{equation} 

\noindent Irrespective of its impact upon RR Lyrae mean periods,
such a change is clearly unrealistic.

\subsection{Period shifts at fixed effective temperature}

\noindent Another way to place constraints on the variations in second-parameter
candidates is to analyze the relative positions of the cluster RR Lyraes on 
the period-effective temperature plane. Of course, the determination of
temperatures for these stars is a rather problematic issue. Caputo \& De
Santis (1992) have advanced a method whereby the ``mass-to-light ratio" of 
RRab Lyrae variables can be obtained in a reddening-independent way,
from periods and blue amplitudes alone. From the period-mean density relation,
in turn, this may be employed to derive RR Lyrae temperatures, and then to
determine period shifts at fixed temperature with respect to the reference
Galactic GC M3. Full details are given elsewhere (Catelan 1996), together
with applications of the method to several Galactic globulars and samples
of field stars.

The $\log P - \log T_{\rm eff}$ diagrams thus obtained for the studied LMC GCs 
are displayed in Fig. 1. The M3 distribution (from Catelan 1996) is reproduced
in the bottom panel. The period shifts were measured with respect to the lower
envelope of the M3 distribution (cf. Catelan 1996), which
is reproduced in all panels for the sake of clarity.

From the definition of period shift, it follows that

\begin{equation}
\Delta \log P (T_{\rm eff}) =
0.84\, \Delta \log L (T_{\rm eff}) - 0.68\, \Delta \log M (T_{\rm eff}).
\end{equation}

\noindent Assuming, as a first approximation, the RR Lyrae masses at a
given temperature in Galactic and LMC globulars to be the same, one finds
that a shift in the RR Lyrae magnitudes by $0.3\, \mbox{mag}$ should actually
imply

\begin{equation}
\delta \Delta \log P (T_{\rm eff}) \approx +0.10.
\end{equation}

\noindent Both Fig. 1 and Table 1 clearly show that such a shift is not allowed
for by the observational data. 

A null LMC -- Galaxy period shift could be produced if the RR
Lyrae masses were higher, in the LMC, by 
$\delta \Delta \log M (T_{\rm eff}) \approx +0.15$ 
[cf. Eq. (9)]. For a mean mass
$\langle M \rangle ^{\rm RR}_{\rm Galaxy} \simeq 0.73\, M_{\sun}$
(as sugested by the Catelan 1993 synthetic
HB models for $Y_{\rm MS} = 0.20$, $Z = 4 \times 10^{-4}$,
$\sigma_{M} = 0.02 \, M_{\sun}$ and not-too-blue HB types), this would
actually demand RR Lyrae masses as high as
$\langle M \rangle ^{\rm RR}_{\rm LMC} \simeq 1.03\, M_{\sun}$ in the 
LMC -- which is probably unrealistic.

\section{Conclusions}

The present analysis of the observational data for RR Lyrae variables in
RR Lyrae-rich LMC GCs suggests that the discrepancy in
the distance moduli of the LMC that are inferred from Cepheid and RR Lyrae
variables cannot be entirely ascribed to variations in second-parameter
candidates. This does {\em not} mean that such variations are not
present, but rather that they would not be sufficient to reconcile
the discrepant distance moduli. A satisfactory explanation of the problem
has yet to be found.

\acknowledgements
The author acknowledges critical readings of the manuscript by 
D. A. VandenBerg, H. J. Rocha-Pinto, and J. E. Horvath.
Financial support by FAPESP is also acknowledged (grant 92/2747-8).

\end{document}